\documentclass[journal,10pt,onecolumn]{IEEEtran}
\usepackage{amsmath}
\usepackage{amsfonts}
\usepackage{amsthm}
\usepackage{dsfont}
\usepackage{epsfig}
\usepackage{upgreek}
\usepackage{subfig}

\usepackage{graphicx}
\graphicspath{{converted_graphics/}}

\usepackage{setspace}

\usepackage{wrapfig}
\usepackage{caption}

\newcommand{\ginv}[1]{{#1}^{\mbox{\tiny -U}}}
\newcommand{\pinv}[1]{{#1}^{\mbox{\tiny -P}}}

\newcommand{\tinv}[1]{\mbox{${#1}^{\overset{\sim}{\mbox{\tiny -1}}}$}}

\begin{document}

\title{A Generalized Matrix Inverse with\\
        Applications to Robotic Systems}       
\author{Bo Zhang and Jeffrey Uhlmann\\
Dept. of Electrical Engineering \& Computer Science\\
University of Missouri-Columbia}
\date{}          
\maketitle

\begin{abstract}
It is well-understood that the robustness of mechanical and robotic control systems
depends critically on minimizing sensitivity to arbitrary application-specific 
details whenever possible.
For example, if a system is defined and performs well in one particular
Euclidean coordinate frame then it should be expected to perform identically 
if that coordinate frame is arbitrarily rotated or scaled. Similarly, the performance
of the system should not be affected if its key parameters are all consistently
defined in metric units or in imperial units. In this paper we show that a 
recently introduced generalized matrix inverse permits performance
consistency to be rigorously guaranteed  in control systems that require
solutions to underdetermined and/or overdetermined systems of equations.\\
~\\
\begin{footnotesize}
\end{footnotesize}
\end{abstract}

\section{Introduction}

Many robotic and mechatronic control systems are mathematically represented,
analyzed, and ultimately implemented as compositions of linear systems.
This is the case even if the dynamics of the system are fundamentally
nonlinear but are solved in terms of linear-algebraic equations or
locally-linear approximations within globally nonlinear state-space models.
In such systems it is commonly necessary to solve an overdetermined
or underdetermined set of equations in order to satisfy a given
set of constraints or to select from a multiplicity of local solutions
within an iterative process. Although this may happen in a single step of
a logically small component of a very large system, if care is
not taken to ensure that critical mathematical properties are
properly preserved the integrity of the overall system can be
compromised. 

It is a fundamental design principal that sensitivity to arbitrary 
application-specific details should be minimized whenever possible.
For example, if a system is defined and performs well in some particular
Euclidean coordinate frame then it should be expected to perform identically 
if that coordinate frame is arbitrarily rotated or scaled. Similarly, the performance
of the system should not be affected if its key parameters are all consistently
defined in metric units or in imperial units. In this paper we show that a 
recently introduced generalized matrix inverse permits performance
consistency to be rigorously guaranteed in control systems that require
solutions to underdetermined and/or overdetermined systems of equations.
Specifically, we show that consistency with respect to arbitrary choices
of units in state-space models of robotic systems can be affected by
a simple replacement of the Moore-Penrose generalized matrix inverse
with a general unit-consistent inverse.

\section{Generalized Matrix Inverses}

For a nonsingular $n\times n$ matrix
$A$ there exists a unique matrix inverse,
$A^{-1}$, which preserves many properties that hold for
ordinary scalar inverses, e.g., matrix inversion distributes
over nonsingular multiplicands as:
\begin{equation}
      (XAY)^{-1} ~ = ~ Y^{-1}A^{-1}X^{-1}
\end{equation}
where noncommutativity of matrix multiplication
imposes a constraint on the ordering of terms but is
otherwise analogous to the scalar case. 
In a practical application the above inverse-distributivity
property implies that if we only have
access to $A$ from its inverse in a linearly
transformed space, $S=(XAY)^{-1}$, then
$A$ can be obtained simply as $YSX$.

When attempting to generalize the notion of a matrix
inverse for singular $A$ it is only possible to 
define an approximate inverse $\tinv{A}$ that retains
a subset of the algebraic properties of a true
matrix inverse \cite{ben}, such as:
\begin{equation}
      A\tinv{A}A ~ = ~ A 
\end{equation}
and
\begin{equation}
      \tinv{A}A\tinv{A} ~ = ~ \tinv{A}
\end{equation}
and/or other properties that may be of analytic
or application-specific utility. The Moore-Penrose  
pseudoinverse~\cite{moore,pen55} (MP inverse), $A^{-P}$, is by 
far the most widely known and used generalized 
inverse\footnote{The Matlab/Octave operator {\sf pinv(M)} 
returns the MP inverse of its matrix argument.}. It is defined 
for any $m\times n$ matrix $A$ and satisfies the above
generalized inverse properties
as well as the following for any conformant 
unitary/orthonormal\footnote{For notational purposes
we retain the generality of interpreting $U$ and
$V$ as arbitrary unitary matrices over $\mathbb{C}$
or $\mathbb{H}$ (e.g., $U^{-1}$ is equal to its
conjugate transpose $U^*$), but
for all practical purposes in this paper they can be 
thought of as representing permutations of 
state variables and/or rotations of a global
coordinate frame.}
matrices $U$ and $V$:
\begin{equation}
    (UAV)^{-P} ~ = ~ V^*A^{-P}U^* \label{udist}
\end{equation}
This property implies that the MP inverse is applicable 
to problems defined in a Euclidean state space for which
the behavior of the system of interest should be invariant
with respect to arbitrary rotations of the coordinate
frame. In that context $A$ can be recovered 
from its MP inverse in a rotationally transformed space as 
$A=V(UAV)^{-P}U$, where consistency
with respect to rigid rotations has been implicitly
exploited. This can be understood by noting that if
$A$ is singular it must be assumed that
$A\neq Y(XAY)^{-P}X$ for arbitrary
nonsingular matrices $X$ and $Y$. More technically,
the MP inverse is consistent with respect to arbitrary unitary
transformations but not to general linear transformations.
 
Despite its widespread default use throughout most areas of 
engineering (often implicitly under the name ``{\em least-squares}''), 
the MP inverse does not satisfy conditions appropriate for
many problems to which it is commonly applied, e.g., ones 
that require consistency with respect to the choice of units
for state variables. For example, a state parameterized with
four variables defined respectively in units relating to temperature, 
pressure, speed, and distance can be thought of as defining a 
4-dimensional Cartesian coordinate frame, but it would make
no sense to rotate that coordinate frame to a space in 
which these variables are mixed. In this
case consistency should be preserved with respect to 
changes of units, e.g., from imperial to metric, rather than
with respect to rotations of a global coordinate frame
which has no physical meaning or interpretation.
This kind of unit consistency (UC)
requires a generalized inverse $\tinv{A}$ that satisfies
\begin{eqnarray}
      \tinv{(DAE)}  & = & E^{-1}\tinv{A}D^{-1}
\end{eqnarray}
where the diagonal matrix $D$ represents units on variables
in one space and the diagonal matrix $E$ represents different 
units for the same variables in a different space.

The hazards associated with the misuse of the
MP inverse have been noted in the robotics 
literature~\cite{duffy90,melch90}, and
disciplined methodologies have been developed to
address the issue in common situations that arise in
that context~\cite{doty93,doty02}. However, the recent derivation 
of an inherently unit-consistent generalized inverse, or UC inverse,
reduces the need for tailored solutions because in
principle it can be simply substituted in place of the MP
inverse~\cite{uhlmann0}. This is not only useful because it simplifies
the implementation process, it also reduces the 
opportunity for subtle implementation errors to
be introduced.

In the next section we examine a robotic system in which we
demonstrate that the MP inverse fails to preserve unit consistency
and thus produces unreliable results. We then show that
simply replacing the MP inverse with the UC inverse provides 
improved and completely stable behavior that is invariant with 
respect to arbitrary changes of units on key parameters.

\section{Generalized Matrix Inverse for Robotics System}

In this section we consider an example involving the motion of a robotic arm. This example is based on systems that have been studied in the literature to show the important aspects of real-world robotics and mechanical system applications~\cite{doty02,huang83}. We will start by describing the structure of the robotic arm and the equations that model its motion so that it can be controlled to perform desired operations. 

\begin{figure}
	\centering
	\includegraphics[width=2.5in]{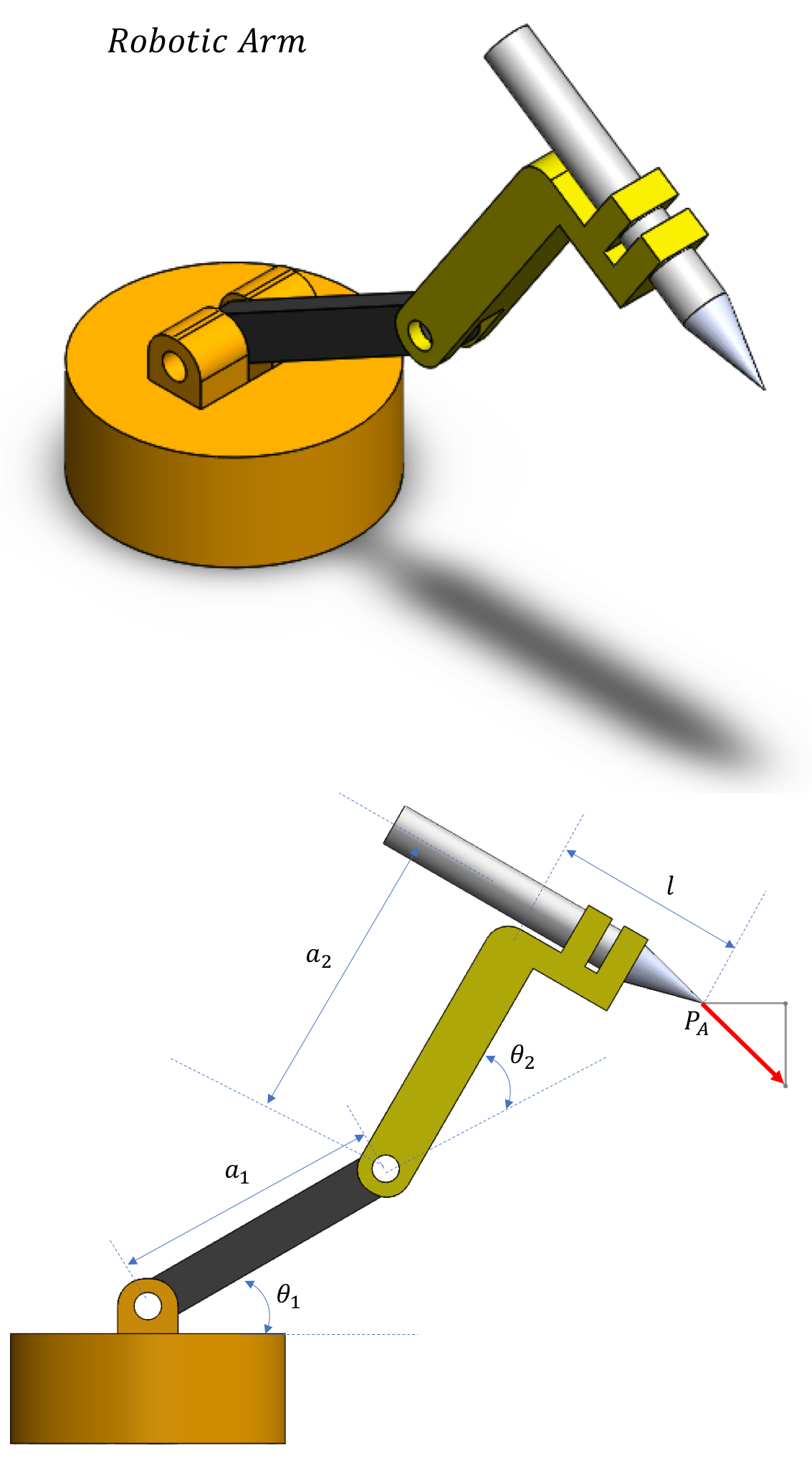}
	\caption[Robotic arm system with two rotational joints and one linear joint.]{Robotic arm system with two rotational joints and one linear joint. The tip-point $P_{A}$ is the end-effector designed to interact with the environment.}	
	\label{f1} 
\end{figure}

The components and configuration of the robotic arm are shown in 
Figure~\ref{f1}. It has structures common in mechanical systems. The tip-point $P_A$ is the end-effector, which denotes the end of a robotic arm and designed to interact with the outside environment. Each joint can be actuated by a motor, and due to the design of the connection, it is limited to 2-dimensional planar motion. The mechanism has 3 degrees of freedom: two of them are rotational and one is linear. These parameters are initialized as $\theta_1 = 30^\circ$, $\theta_2 = 30^\circ$ and $l = 0.7 m$. A mixed control of these joints will determine the motion of tip-point $P_A$, e.g., to take it to a desired state B. Taking the cylindrical joint on the frame as the origin, the position of the tip-point $P_A$($[x, y, z]$) is given by equations \ref{3-6}-\ref{3-8}.
\begin{eqnarray}
	x &=& a_{1}c_{1} + a_{2}c_{12} + l \cdot s_{12} \label{3-6} \\
	y &=& a_{1}s_{1} + a_{2}s_{12} - l \cdot c_{12} \label{3-7} \\
	z &=& 0 \label{3-8} 
\end{eqnarray}
where $s_{1} = \sin(\theta_1), c_{1} = \cos(\theta_1), s_{12} = \sin(\theta_1 + \theta_2), c_{12} = \cos(\theta_1 + \theta_2)$. A Jacobian matrix represents the transformation of the end-effector representation of 
the dynamic system ($P_A$) into a joint-state representation in which all of the key parameters are represented as a vector as $q$. For the current problem the state of the system is given by
$q = [\theta_{1} , \theta_{2}, l]^T$, which are the states of the 3 joints. The Jacobian matrix for this system,
\begin{equation} \label{3-9}
	J = \frac{\partial P_A}{\partial q} 
\end{equation}	
is determined to be\\	
$$
\begin{bmatrix}
	-a_{1}s_{1} - a_{2}s_{12} + l \cdot c_{12}, & - a_{2}s_{12} + l \cdot c_{12}, & s_{12} \\
	a_{1}c_{1} + a_{2}c_{12} + l \cdot s_{12}, & a_{2}c_{12} + l \cdot s_{12}, & - c_{12} \\
	0,      &     0,    & 0 
\end{bmatrix}
$$
The target velocity is thus
\begin{eqnarray}
	J = \frac{\partial P_A}{\partial q} &=& \frac{\partial P_A}{\partial t} \cdot \frac{\partial t}{\partial q} \label{3-10} \\ 
	\frac{\partial P_A}{\partial t} &=& J \frac{\partial q}{\partial t} \label{3-11} 
\end{eqnarray}
or
\begin{equation} \label{3-12}
	\vec{v} = J\dot q
\end{equation}
where $\vec{v}$ is the derivative of $P_A$, and $\dot q$ is the joint velocity of the robotic arm represented by $\dot q = 
[\dot \theta_{1} , \dot \theta_{2}, \dot l]^T$. Assuming a target velocity $\vec{v} = [2, -2, 0]^T (m/s )$, the solution to achieve the desired motion is
\begin{equation} \label{3-13} 
	\dot q = J^{-1}\vec{v} 
\end{equation}
but this cannot be evaluated if $J$ is singular, {\em and it 
will be singular} in this case because the motion is constrained 
to a 2-dimensional plane in a 3-dimensional space. The solution
therefore requires use of a generalized 
matrix inverse in place of the undefined matrix inverse. 
As has been discussed, the most commonly used generalized 
inverse is the Moore-Penrose (MP) inverse, which gives a result
for $\dot q$ as
\begin{equation} \label{3-14}
	\dot q = J^{-P}\vec{v} 
\end{equation}
where $J^{-P}$ is the MP inverse. As will be seen, the 
use of the MP inverse is not appropriate if the state
vector has variables with units that must
be preserved, which is true for the case of $q$. In this system lengths are defined in meters, $a_{1} = 1 m$, $a_{2} = 1.1 m$, $l = 0.7 m$, and $\vec{v} = [2, -2, 0]^T (m/s )$, 
though the system should be expected to work correctly no matter what units are used as long as everything is defined consistently in those units.  

\begin{figure}
	\centering
	\includegraphics[width=3.5in]{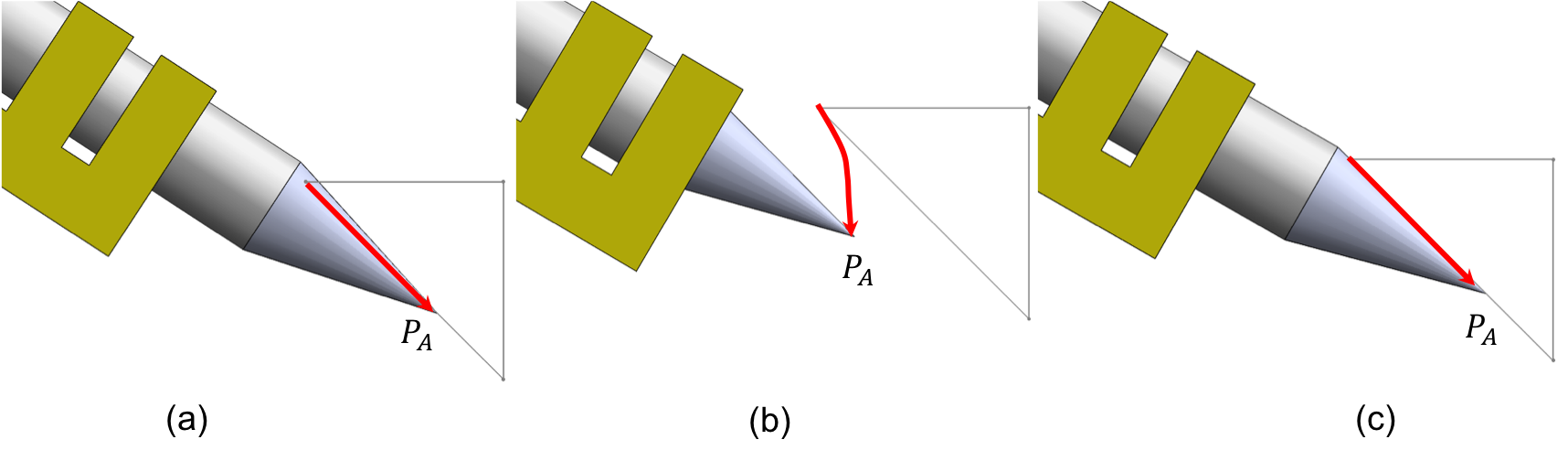}
	\caption{Tip point position after simulation, solved with MP inverse. 
	(a) length unit(m), timestep ($10^{-3} s$).
	(b) length unit(cm), timestep ($10^{-3} s$). (c) length unit(cm), timestep ($10^{-4} s$). Simulation (b) fails by deviating from designed track.}	
	\label{f2} 
\end{figure}

The joint velocity needed
to keep the speed of tip-point $P_A$ equal to $\vec{v}$ is calculated using equation \ref{3-14} with timesteps of $dt = 10^{-3} s$ over a simulation time of 0.1 s. At each timestep (e.g., $t = 0s, 
t = 10^{-3}, t = 2\times 10^{-3}$)  the angles and lengths are updated for the next iteration as $q^{(1)} = q^{(0)} + \dot q^{(0)} \cdot dt$ so that the Jacobian matrix can also be updated 
to solve the $\dot q$ for the next timestep. The calculated velocities included in Table \ref{t1} can be seen to   
change slowly over time, and the simulation shows that they lead to a tip-point velocity that is approximately equal to $\vec{v}$ during the process. The final state of the tip-point is shown in Figure \ref{f2}(a), 
which shows that the target was successfully controlled to move along the designed track.  
 \begin{table}[!t]
\renewcommand{\arraystretch}{1.3}
\caption{Joint velocity of robotic arm calculated using MP inverse. 
	Timestep: 0.001s.}
\label{t1} 
	\centering \begin{tabular}{|r||l|l|l|l|}
	\hline
	Time($s$) & $\dot \theta_{1}(degree / s)$ & $\dot \theta_{2}(degree / s)$ & $ \dot l(m / s)$ & $V^T(m / s)$ \\ 
	\hline
	0.000 & -27.881 & -12.12 & -1.543 & [2, -2, 0]\\
	\hline
	0.001 & -27.826 & -11.981 & -1.548 & [2, -2, 0]\\
	\hline
	0.002 & -27.772 & -11.838 & -1.553 & [2, -2, 0]\\
	\hline
	0.003 & -27.719 & -11.695 & -1.558 & [2, -2, 0]\\
	\hline
	0.004 & -27.666 & -11.552 & -1.563 & [2, -2, 0]\\
	\hline
	0.005 & -27.614 & -11.409 & -1.568 & [2, -2, 0]\\
	\hline
	0.006 & -27.563 & -11.266 & -1.573 & [2, -2, 0]\\
	\hline
	0.007 & -27.513 & -11.123 & -1.578 & [2, -2, 0]\\
	\hline
	0.008 & -27.464 & -10.980 & -1.582 & [2, -2, 0]\\
	\hline
	0.009 & -27.414 & -10.837 & -1.587 & [2, -2, 0]\\
	\hline
	0.010 & -27.367 & -10.693 & -1.592 & [2, -2, 0]\\
	\hline
	\end{tabular}
	\end{table}
\vspace*{11pt}

In contrast to the first simulation, the second simulation is performed identically except that lengths are defined in units of centimeters instead of meters: $a_{1} = 100 cm$, $a_{2} = 110 cm$, $l = 70 cm$, and $\vec{v} = [200, -200, 0]^T (cm/s )$. As a result, some elements of the Jacobian matrix are changed in magnitude, but the overall behavior of the system should be expected to remain unchanged. In other words, as long as everything is implemented consistently using lengths defined in meters, or implemented consistently using lengths defined in centimeters, the behavior of the system should be the same because the controls will be computed accordingly. The choice of units should not matter, and if it does then the system cannot be trusted. 

\begin{figure} 
	\centering
	\subfloat[]{\includegraphics[width=2.5in]{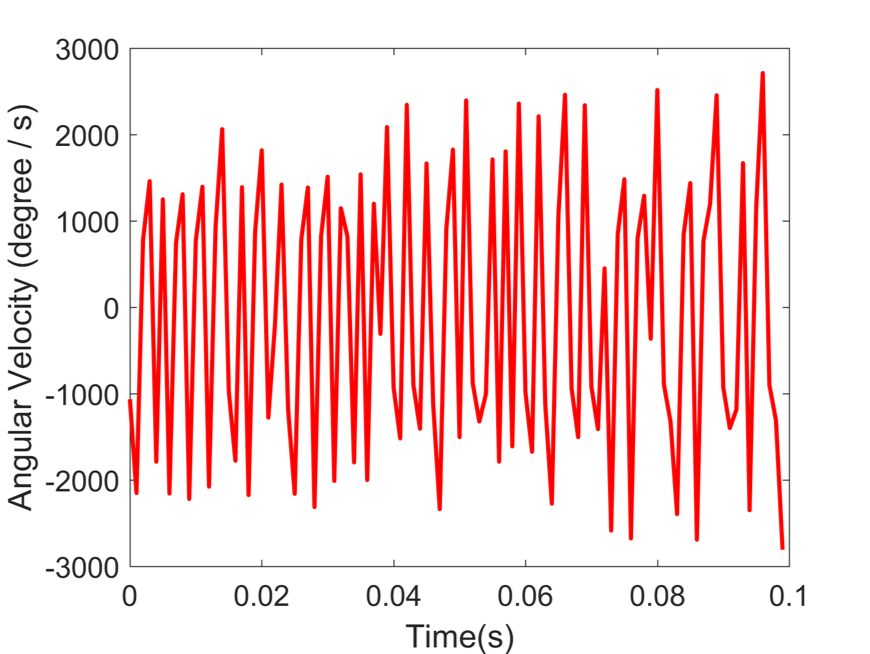}%
	\label{f3-1}}
	\hfil
	\subfloat[]{\includegraphics[width=2.5in]{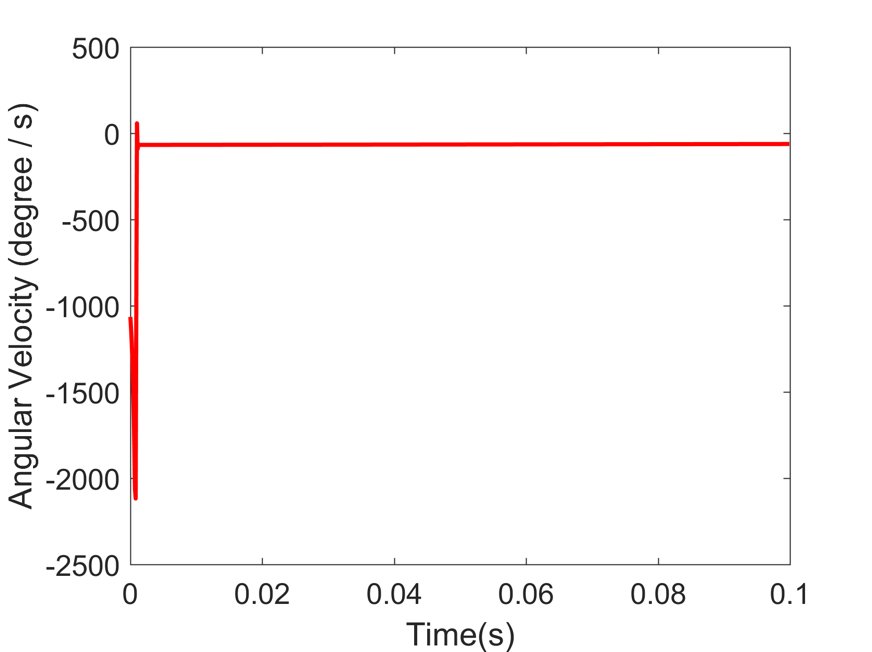}%
	\label{f3-2}}
	\hfil
	\subfloat[]{\includegraphics[width=2.5in]{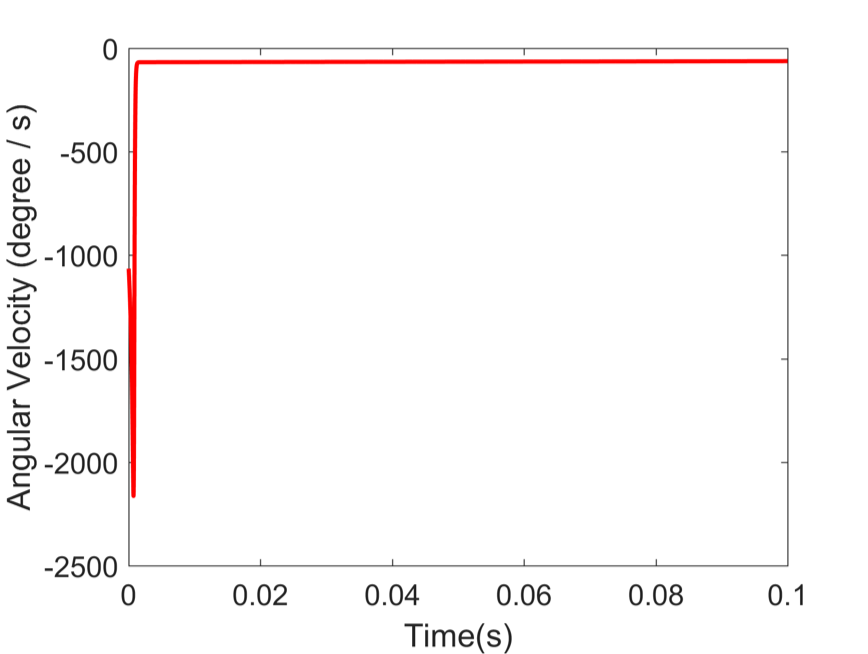}%
	\label{f3-3}}
	\hfil
	\subfloat[]{\includegraphics[width=2.5in]{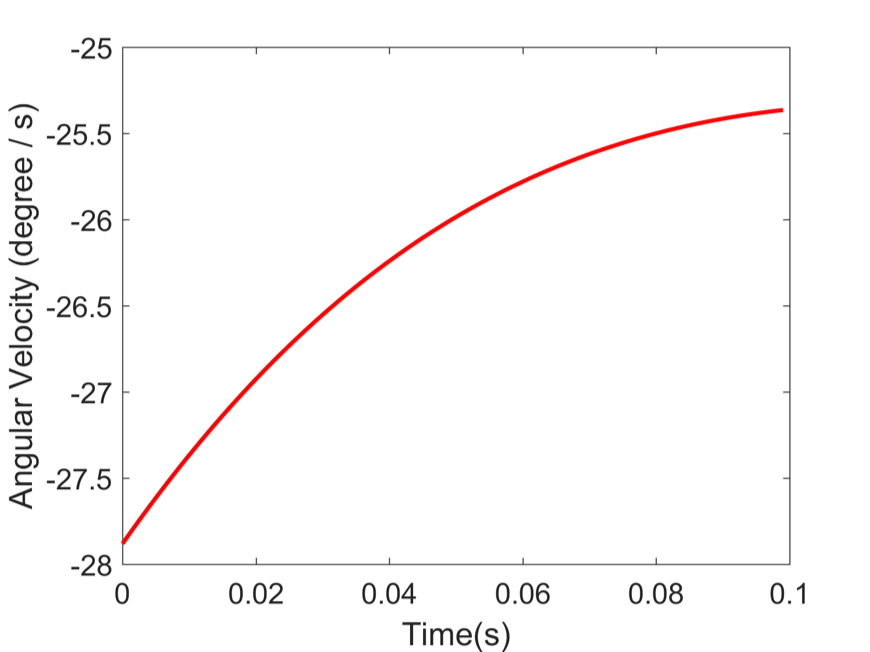}%
	\label{f3-4}}
	\caption{Joint velocity ($\dot \theta_{1}$) solved with MP inverse. 
	(a) length unit(cm), timestep ($10^{-3} s$).
	(b) length unit(cm), timestep ($10^{-4} s$). (c) length unit(cm), timestep ($10^{-5} s$). 
	(d) length unit(m), timestep ($10^{-3} s$). Simulation (a) does not converge, and simulation (b) and (c) can converge with a smaller timestep.}	
	\label{f3}
\end{figure}

Although the joint velocities are calculated the same way as before with a timestep of $10^{-3} s$, the system now diverges rapidly. For example, Figure \ref{f3}{(a)} shows that the value of the angular velocity $\theta_1$ 
rapidly fluctuates between $[-2000, 2000]~rad/s$, whereas the computed angular velocity in the previous simulation 
remained stable at around $30~rad/s$. Figure \ref{f2}{(b)} shows that the control is unsuccessful as the tip-point motion diverges from the designed path. 
To reduce the magnitude of this deviation the control system has to be performed 
using much smaller timesteps, which requires control operations to be calculated and 
applied much more frequently to keep the tip from diverging too far from the correct path.

The results of several tests are plotted in Figure \ref{f3}. It can be seen that
acceptable results are obtained with a timestep of $10^{-3} s$ when lengths are defined in units of meters, but unstable results are produced
when the length unit is changed to centimeters. In the centimeter case it is necessary to reduce the timestep by an order of magnitude to $10^{-4} s$ to achieve reliable control, but this increases the computational complexity by an order of magnitude because control operations must be calculated much more frequently. While this does yield an acceptable tip-point velocity, the control values are different from those produced when units were in meters and this is evidence that something is wrong. It is tempting to conclude based
on the final state displayed in Figure \ref{f2}(c) that the controls perform equally well in both cases because they produce seemingly similar paths. However, Figure \ref{f3}(a) shows that the control produced in the case of centimeter units leads to very erratic high-frequency motion along the path, and the controls required to achieve these wildly changing velocities may be physically difficult to actually achieve with real hardware. Even if they can be achieved they may cause excessive stress and wear on machine elements. 

The simulations clearly show that different units lead to different configurations for the same problem when using the MP inverse. The final state of the robotic arm is $\theta_1 = 27.379^\circ$, 
$\theta_2 = 29.483^\circ$ and $l = 0.875 m$ for lengths in meters, $\theta_1 = 22.109^\circ$, $\theta_2 = 38.129^\circ$ and $l = 0.864 m$ for lengths in centimeters. The fact that results depend on the choice of units raises concerns because there is no way to predict when and how this dependency may lead to bad results. Although the timestep can be reduced until the controls calculated using the MP inverse produce acceptable results, the need to adjust the timestep to compensate for arbitrary dependencies on the choice of units implies that the solution has been tailored to the problem and the results may not remain acceptable for a slightly different configuration. Therefore, this control system using the MP inverse cannot be trusted.

It is important to review why the control system in this case seemed to work well when units were defined in meters but not when
they were defined in centimeters. This happened because the control calculations used the MP inverse when there was need to invert
a singular Jacobian matrix. Based on the properties discussed in the previous chapter, the MP inverse is defined to provide a
solution to a linear system $Ax = y$ that minimizes the Euclidean norm $||Ax - y||_2$ for an over-constrained system, or to minimize the norm $||x||_2$ for an under-constrained system. 
While this makes sense for vectors in which elements are all defined in the same Euclidean coordinate system, it does not make sense 
when the elements (parameters) are not defined in a common coordinate frame and have incommensurate units. 
This is because the choice of units affects the magnitude of ``error'' that is contributed by 
each element when the sum of squared errors is minimized. For our case the MP inverse attempts to minimize 
\begin{equation} \label{3-15} 
	||\dot q||_2 = \dot \theta_{1}^2(degree^2 / s^2) + \dot \theta_{2}^2(degree^2 / s^2) + \dot l^2 (m^2/s^2) 
\end{equation}
where the relative contributions of different terms clearly depend on the arbitrary choices of units for the different elements.
This shows that squared-error is not a physically meaningful quantity to minimize when different variables are defined in different units. 
The loss of physical meaning when minimizing Euclidean inner products involving variables with incommensurate units has been observed
to be problematic for robotic systems \cite{duffy90,doty93}, and essentially the same problem applies in our case when the MP inverse is
applied to the Jacobian matrix representing the transformation of variables in incommensurate units. This is why our simulation results
were not consistent with respect to changes of units from meters to centimeters. The fact that minimizing squared error has no 
physical meaning in this context
reinforces the conclusion that the calculated controls using the MP inverse should not be trusted.

At this point it is valuable to consider what property of a generalized inverse is needed to avoid the unit dependency problems
of the MP inverse. What is needed is a solution that is consistent with respect changes of units in the governing equation $Ax = y$, 
which can be expressed in the form of diagonal matrices $D$ and $E$ as
\begin{eqnarray}
	x' = Dx \label{3-16} \\
	y' = Ey \label{3-17}
\end{eqnarray}
where $x'$ is just $x$ but with elements defined in different units and $y'$ is similarly obtained from $y$.
The governing equation can now be expressed as
\begin{equation} \label{3-18}
	y' = EAx = EAD^{-1}x' 
\end{equation}
The solution can be expressed in terms of an unknown generalized inverse $(EAD^{-1})^{\widetilde{-1}}$ as
\begin{equation} \label{3-19}
	x' = EAx = (EAD^{-1})^{\widetilde{-1}}y'
\end{equation}
while the solution in the original units was $x = A^{\widetilde{-1}}y = A^{\widetilde{-1}} E^{-1}y'$. In order to satisfy our assumption
that changes of units take the form of diagonal transformations, e.g., $x' = Dx$ and $y' = Ey$ , the unknown general inverse must satisfy 
\begin{equation} \label{3-20}
	(EAD^{-1})^{\widetilde{-1}}  = DA^{\widetilde{-1}} E^{-1}
\end{equation}
which is not satisfied by the MP inverse because it only provides consistency with respect to orthogonal transformations (rotations) but not diagonal transformations.
The problem to be solved requires a generalized inverse that ensures consistency for diagonal transformations, and an inverse of this kind
has been recently developed \cite{uhlmann}. This unit-consistent (UC) inverse can be defined in terms of the MP inverse as follows:
\begin{eqnarray}
	A^{-U} &=&  (D_ASE_A)^{-U} \label{3-21} \\
	&=& E_A^{-1}S^{-P}D_A^{-1} \label{3-22}
\end{eqnarray}
where $D_A$ and $E_A$ are positive diagonal matrices determined from $A$, and $S$ is a matrix uniquely determined from $A$ so that the magnitude of the product 
of the nonzero elements of each row and column of $S$ is 1. (The basis for this decomposition, and its existence and uniqueness properties, are
described in~\cite{uhlmann}.) 
The UC inverse works by eliminating the effect of diagonal transformations. From the definition it can be seen that
\begin{eqnarray}
	A^{-U} &=& (D_A^{-1}D_AAE_AE_A^{-1})^{-U} \label{3-23}\\
	&=& E_A(D_AAE_A)^{-P}D_A \label{3-24}
\end{eqnarray}
So for any diagonal matrices $D$ and $E$ applied to change the units,
\begin{equation} \label{3-25}
	(EAD^{-1})^{-U} = (ED_A^{-1}D_AAE_AE_A^{-1}D^{-1})^{-U} 
\end{equation}
and since $E_A^{-1}D^{-1}$ and $ED_A^{-1}$ are also diagonal matrices: 
\begin{eqnarray}
	&&(EAD^{-1})^{-U} \nonumber \\
	&=& (E_A^{-1}D^{-1})^{-1}(D_AAE_A)^{-P}(ED_A^{-1})^{-1} \label{3-26}\\
	&=& DE_A(D_AAE_A)^{-P}D_AE^{-1} \label{3-27}\\
	&=& DA^{-U}E^{-1} . \label{3-28} 
\end{eqnarray}
This shows that the UC inverse satisfies the consistency requirement of equation \ref{3-20}.  

The problems we have demonstrated with the MP inverse occur because it does not produce control behavior that is invariant with respect to the choice of units. Assuming the motion is given as $\vec{v}_m = J_m \dot q_m$ in units of meters, with the Jacobian matrix terms given in equation \ref{3-9} and $\vec{v}_m = [\vec{v}_{x} , \vec{v}_{y} , \vec{v}_{z} ]^T$, the transformation of units from
meters to centimeters, $\vec{v}_{cm} = J_{cm} \dot q_{cm}$, can be expressed as a diagonal transformation in the form of:
\begin{equation}
	\begin{bmatrix}
	c, & 0, & 0 \\
	0, & c, & 0 \\
	0,  & 0, & c 
	\end{bmatrix} \cdot \vec{v}_m = J_m \cdot
	\begin{bmatrix}
	c, & 0, & 0 \\
	0, & c, & 0 \\
	0, & 0, & 1 
	\end{bmatrix} \cdot  \dot q_{cm} \label{3-29}
\end{equation}
where $c=100$ is the scale factor of converting from meters to centimeters. What is needed is a generalized inverse that will guarantee invariance with respect to units. The UC inverse has this property, so we should expect that using it in place of the MP inverse will eliminate unit dependencies in the control calculations and avoid the unstable behavior demonstrated in our earlier simulations. This can be tested simply by re-running the simulations with the MP inverse replaced by the UC inverse and checking whether the performance of the control system is invariant with respect to changes of units from meters to centimeters. Said another way, changing the units should produce a control solution defined in the new units but which exhibits the exact same physical behavior obtained using the original units.

\begin{figure} 
	\centering
	\includegraphics[width=3in]{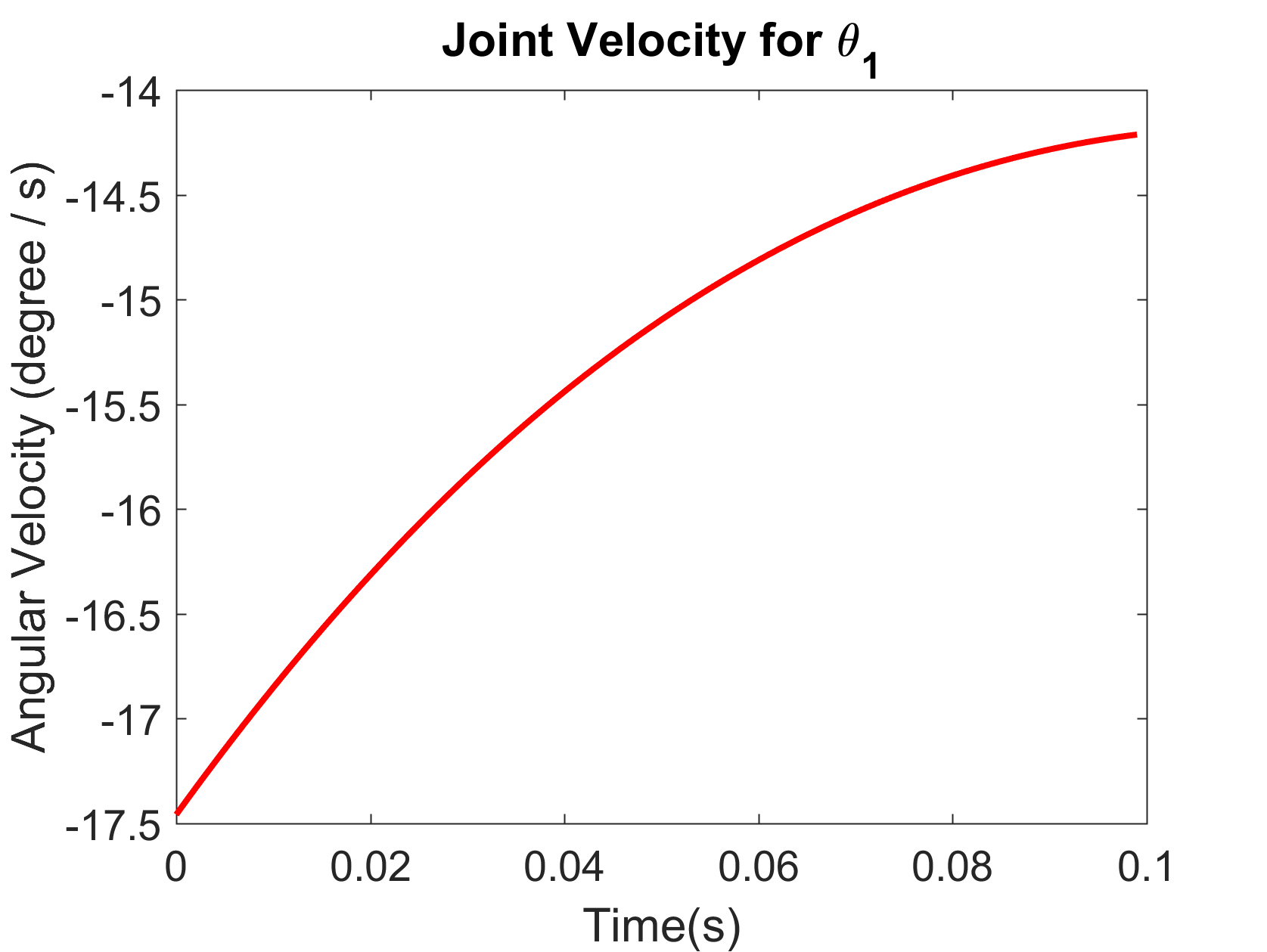}
	\caption{Joint velocity ($\dot \theta_{1}$) generated by simulations using both length units(meter / centimeter) and timestep ($10^{-3} s$). The result changes smoothly over the simulation time.}	
	\label{f4} 
\end{figure}

As was done in the earlier simulations, controls are calculated with a timestep of $10^{-3} s$ over a simulation time of $0.1s$ for the case of lengths defined in meters 
and then with lengths in centimeters according to equation \ref{3-14}. Figure \ref{f4} shows that in both cases the control produced using the UC inverse converges rapidly, 
and the velocity variation is identically smooth in both cases.  This corroborates our expectation that the UC inverse produces stable results that are not affected by the arbitrary choice of units used for lengths. 

It should be noted that we showed in the earlier simulations that the timestep could be reduced so that controls from the MP inverse also produced acceptable results. The difference is that the UC-inverse solution mathematically guarantees that its results are invariant with respect to the choice of units and therefore does not require timestep changes to avoid unpredictable behaviors. The MP inverse does not have the correct mathematical properties and therefore cannot offer the same guarantee. 

\section{Generalized Consistency Considerations}

In the previous section it was shown that the UC inverse satisfied all requirements 
for maintaining unit consistency in the example system\footnote{An additional
property that is sometimes required of a generalized inverse that was not
demonstrated in our example system is consistency with respect to
use of the Kronecker product. Because this property was not among
those established in~\cite{uhlmann}, we formally prove it in
Appendix~A.}. In this secton we examine how the UC inverse can also be combined 
with the MP inverse, and even other generalized inverses (e.g., the Drazin
inverse~\cite{drazin,cmr76,drazinApp5} or other similarity-consistent inverse~\cite{jkusim}), to construct solutions to inverse problems when 
there is a mix of variables involving different consistency requirements. In addition to variables that require unit consistency to be preserved, a complex real-world system may also involve variables defined in a Cartesian coordinate frame that require consistency with respect to rotations of that coordinate frame. In other words, the behavior of the control system must be invariant with respect to changes of units for some variables and invariant with respect to rotations for other variables. The UC inverse is applicable in one case while the MP inverse is applicable in the other, but what is needed for such a system is a generalized inverse that will guarantee unit consistency for some variables and rotation consistency for others. If we assume\footnote{The ordering of the variables is arbitrary so there is no loss of generality in assuming they are permuted so that the UC variables come first and the rotation-consistent variables come next.} that the first $m$ variables require unit consistency and the remaining $n$ variables require rotation consistency then the transformation matrix to be inverted can be block-partitioned as
\begin{eqnarray}
   \begin{array}{rcl}
      A & = &
         \left[ \begin{array}{cc} { W} & { X}\\
                             { Y}& { Z}\end{array} \right]
                       \begin{array}{l} { \}}~m \\{ \}}~n \end{array}\\
                        & &  
                                \begin{array}{c}                                                          
                                   \;\underbrace{\;}_m\;\underbrace{\;}_n
                                \end{array}
   \end{array} \label{4-37}
\end{eqnarray}
It has been shown \cite{uhlmann} that the mixed inverse can be obtained from this block-partitioned form as
\begin{equation}
   A^{-M} = \nonumber 
\end{equation}
\begin{equation}
   \left[
     \begin{array}{cc}
     \ginv{({ W}-{ X}\pinv{{ Z}}{ Y})} & -\ginv{{ W}}{ X}\pinv{({ Z}-{ Y}\ginv{{ W}}{ X})} \\
      -\pinv{{ Z}}{ Y}\ginv{({ W}-{ X}\pinv{{ Z}}{ Y})} & \pinv{({ Z}-{ Y}\ginv{{ W}}{ X})} 
     \end{array}
    \right] \label{4-38}
\end{equation}

In the previous examples it has been stressed that the problem solution should not depend on arbitrary system preferences because such dependencies lead to unpredictable results. In the case of a system with a mix of consistency requirements we should expect the correct solution to produce the same behavior if units on the first $m$ variables are changed or if the coordinate frame of the remaining $n$ variables is rotated. If the mixed inverse above is correct then the controls produced should have this property. We will demonstrate that it does in the following example.

\begin{figure} 
	\centering
	\subfloat[]{\includegraphics[width=2.5in]{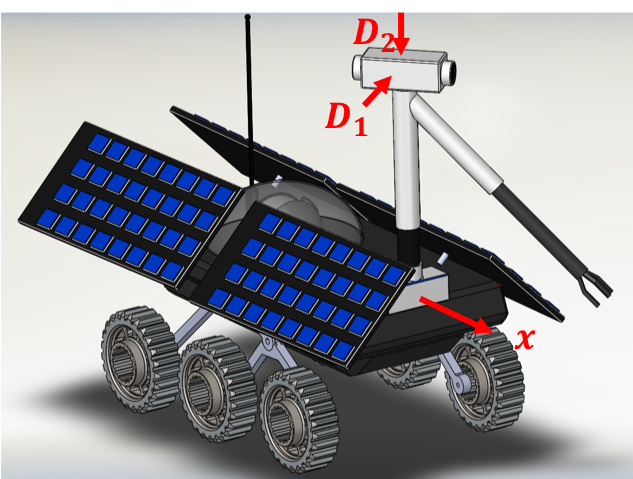}%
	\label{f5-1}}
	\hfil
	\subfloat[]{\includegraphics[width=2.5in]{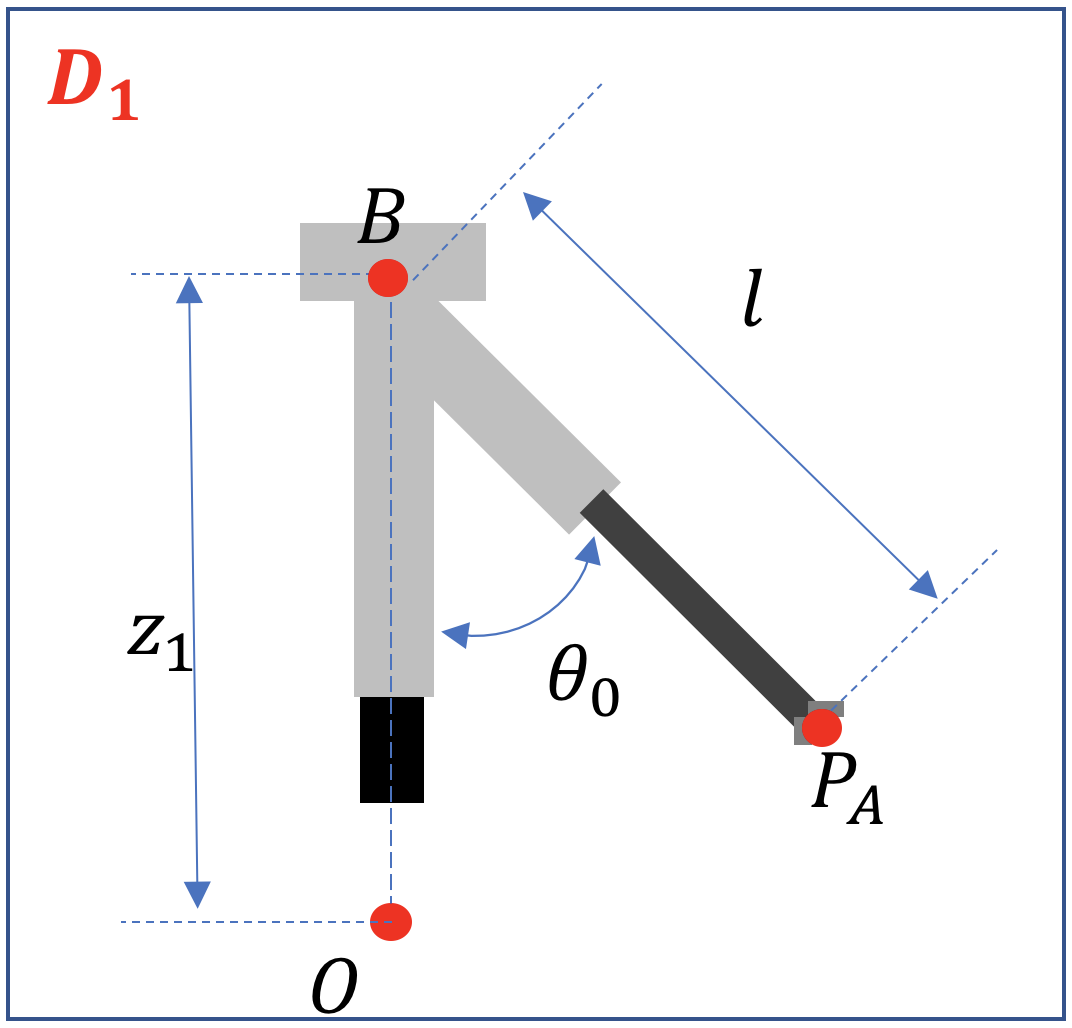}%
	\label{f5-2}}
	\hfil
	\subfloat[]{\includegraphics[width=2.5in]{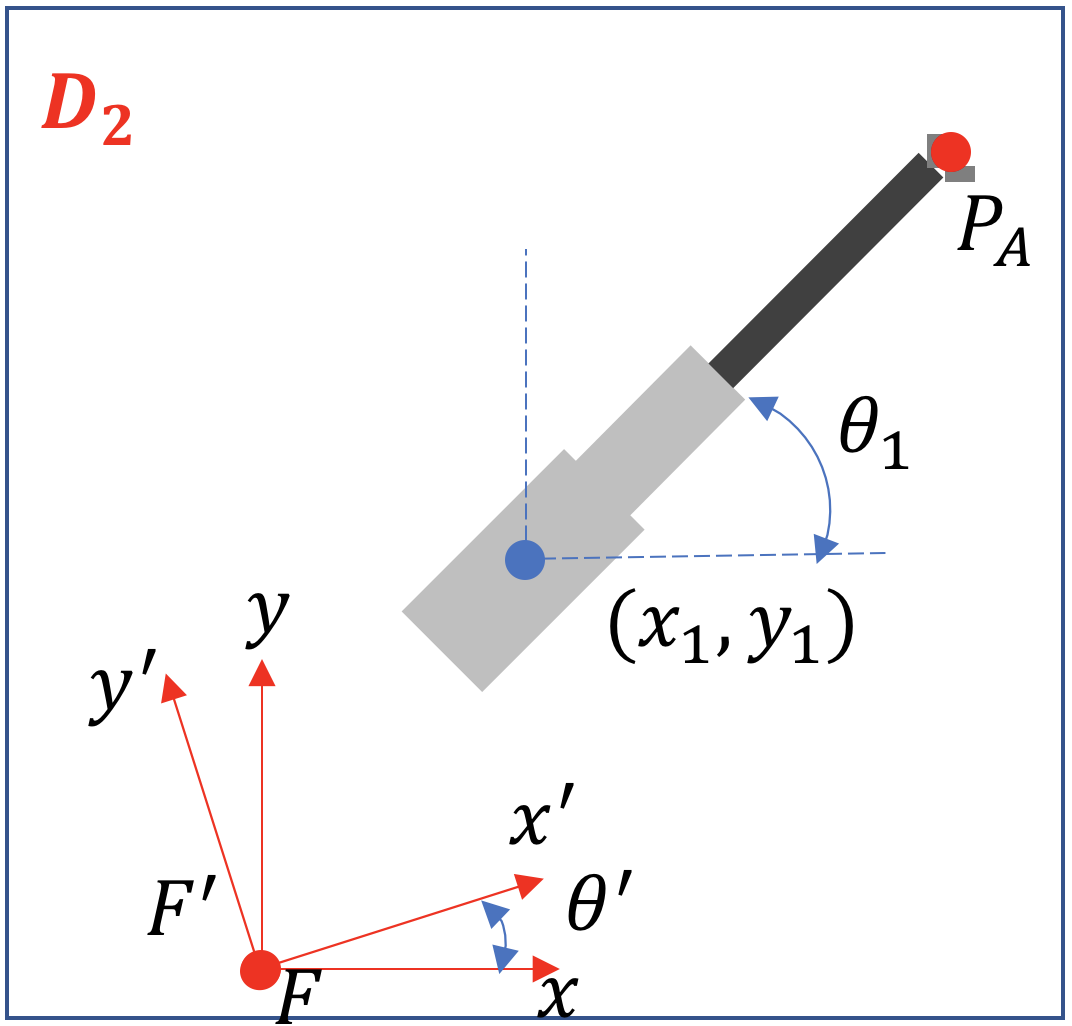}%
	\label{f5-3}}
	\caption[]{Rover with an extendable arm. The rover is free to move on a plane, body part B can ascend/descend or rotate, and the arm can be extended. The two projected views, $D_1$ and $D_2$, show the structure of the arm. $\theta_0 = 45^\circ$ is a fixed angle. The coordinate frame $F'$ is a rotation of the original frame $F$ by an angle of $\theta'$.}	
	\label{f5} 
\end{figure}

Consider the control of a planetary rover with a robotic arm as displayed in Figure \ref{f5}, which includes two projected views in directions $D_1$ and $D_2$. The x-y coordinate is shown as frame $F$. The frame $F'$, which is an orthogonal transpose of frame $F$ by $\theta'$, will be considered later. The rover is free to move in any direction on planar terrain, and its Cartesian position coordinates in this plane are $(x_1, y_1)$. The part B can rotate and can also ascend/descend\footnote{The part B can be thought of as an extendible arm for taking a soil or rock sample. This rover, called ODIF, is intended to function as part of a team of lightweight bots for large area investigation.}. In addition, the arm can elongate within a fixed range, but it cannot rotate in the vertical plane. Thus there are 5 degrees of freedom for the design, denoted as $q = [\theta_1, l, x_1, y_1, z_1]^T$, where $(x_1, y_1, z_1)$ is the position coordinates of part B. 

From the geometry relations, the position of tip-point $P_A$ is given as $P_A = [x, y, z, 0, 0]^T$,
 \begin{eqnarray} 
	x &=& x_1 + l \cdot \sin \theta_0 \cdot \cos \theta_1 \label{4-39}\\
	y &=& y_1 + l \cdot \sin \theta_0 \cdot \sin \theta_1 \label{4-40}\\
	z &=& z_1 - l \cdot \cos \theta_0 \label{4-41}
\end{eqnarray}
We can use the same approach of the previous example to generate the Jacobian matrix for $\vec{v} = J\dot q$ as 
\begin{center}
$$
J = 
\begin{bmatrix}
	-l\cdot \sin \theta_{0}\sin\theta_{1} & \sin\theta_{0}\cos\theta_{1} & 1 & 0 & 0 \\
	l\cdot \sin \theta_{0}\cos\theta_{1} & \sin\theta_{0}\sin\theta_{1} & 0 & 1 & 0 \\
	0 & -\cos\theta_{0} & 0 & 0 & 1 \\
	0 & 0 & 0 & 0 & 0 \\
	0 & 0 & 0 & 0 & 0
\end{bmatrix}
$$
\end{center}

The initial states are set to be $\theta_1 = 45^\circ$ and $l = 1.0 m$, the constant angle $\theta_0 = 45^\circ$. The target velocity of the tip-point is $\vec{v} = [2, 0, -1, 0, 0] m/s$. Since the system 
has redundant degrees of freedom and therefore $J$ is singular, $\dot q = J^{\widetilde{-1}}\vec{v}$ has to be solved with a general inverse (e.g. MP inverse, UC inverse, or mixed inverse).
 For the unknown variables in $\dot q$, $[\dot \theta_1, \dot l]$ have incommensurate units, and $[\dot x_1, \dot y_1, \dot z_1]$ are defined in a common Euclidean space.
  Thus the Jacobian matrix can be partitioned as
$$
W = 
\begin{bmatrix}
	-l\cdot \sin \theta_{0}\sin\theta_{1} & \sin\theta_{0}\cos\theta_{1}  \\
	l\cdot \sin \theta_{0}\cos\theta_{1} & \sin\theta_{0}\sin\theta_{1} \\
\end{bmatrix},
X = 
\begin{bmatrix}
	1 & 0 & 0 \\
	0 & 1 & 0
\end{bmatrix} 
$$
$$
Y = 
\begin{bmatrix}
	0 & -\cos\theta_{0} \\
	0 & 0 \\
	0 & 0
\end{bmatrix},
Z = 
\begin{bmatrix}
	0 & 0 & 1 \\
	0 & 0 & 0\\
	0 & 0 & 0
\end{bmatrix} 
$$

Now the mixed inverse can be used to find the solution for $\dot q$ using equation \ref{4-38}. We initially use meters as the length unit and $F$ as the coordinate frame. We will then consider a change of length units from meters to centimeters and a coordinate frame rotation from $F$ to $F'$ by a rotation angle of $\theta' = 30^\circ$. Given $c=100$ as the scale factor to convert from meters to centimeters, the governing equation for the centimeter and rotated case, $\vec{v}_{cm, F'} = J_{cm, F'} \dot q_{cm, F'}$, can be expressed as a diagonal transformation of the meter case, $\vec{v}_{m, F} = J_{m, F} \dot q_{m, F}$, as
\begin{eqnarray*}
	&&
	\begin{bmatrix}
	c\cos(\theta'), & -c\sin(\theta'), & 0, & 0, & 0 \\
	c \sin(\theta'), & c\cos(\theta'), & 0, & 0, & 0 \\
	0, & 0, & c, & 0, & 0 \\
	0, & 0, & 0, & c, & 0\\
	0, & 0, & 0, & 0, & c
	\end{bmatrix} \cdot \vec{v}_{m, F} \\ \label{4-42}
	&=&\begin{bmatrix}
	 \cos(\theta'), & -\sin(\theta'), & 0, & 0, & 0 \\
	\sin(\theta'), & \cos(\theta'), & 0, & 0, & 0 \\
	0, & 0, & 1, & 0, & 0 \\
	0, & 0, & 0, & 1, & 0\\
	0, & 0, & 0, & 0, & 1
	\end{bmatrix} \cdot
	J_{m, F} \\ &&\cdot
	\begin{bmatrix}
	c, & 0, & 0, & 0, & 0 \\
	0, & 1, & 0, & 0, & 0 \\
	0, & 0, & 1, & 0, & 0 \\
	0, & 0, & 0, & 1, & 0\\
	0, & 0, & 0, & 0, & 1
	\end{bmatrix} \cdot  \dot q_{cm, F'} \label{4-43}
\end{eqnarray*}
This shows the block of the matrix requiring rotation consistency and the block of variables defined in incommensurate units. We then test the three possible approaches
to computing the controls: using the MP inverse alone; using the UC inverse alone; and using the mixed inverse obtained from equation \ref{4-38}.
The solutions for $\dot q$ from the three approaches are displayed in table \ref{t2} for $t = 0s$. The column headings give the unit/coordinate frame in which each test was performed but the results are all given in (converted to) a common coordinate frame for comparison purposes. As can be seen, the mixed inverse is the only approach that produces identical results regardless of coordinate-system changes. For the other approaches, it can be seen that when the controls are evaluated with the MP inverse, it generates the same results when different coordinate frames are used but not when length units are changed. When the UC inverse is solely applied to the entire system it generates invariant solutions when units are changed but not when rotations are applied. Therefore, it can be concluded that solely using either the UC inverse or MP inverse alone will not produce reliable results. Instead, the mixed inverse is required to ensure that the behavior of the system is invariant with respect to defined changes of units and coordinates. 

\begin{figure}
	\centering
	\subfloat[]{\includegraphics[width=3.3in]{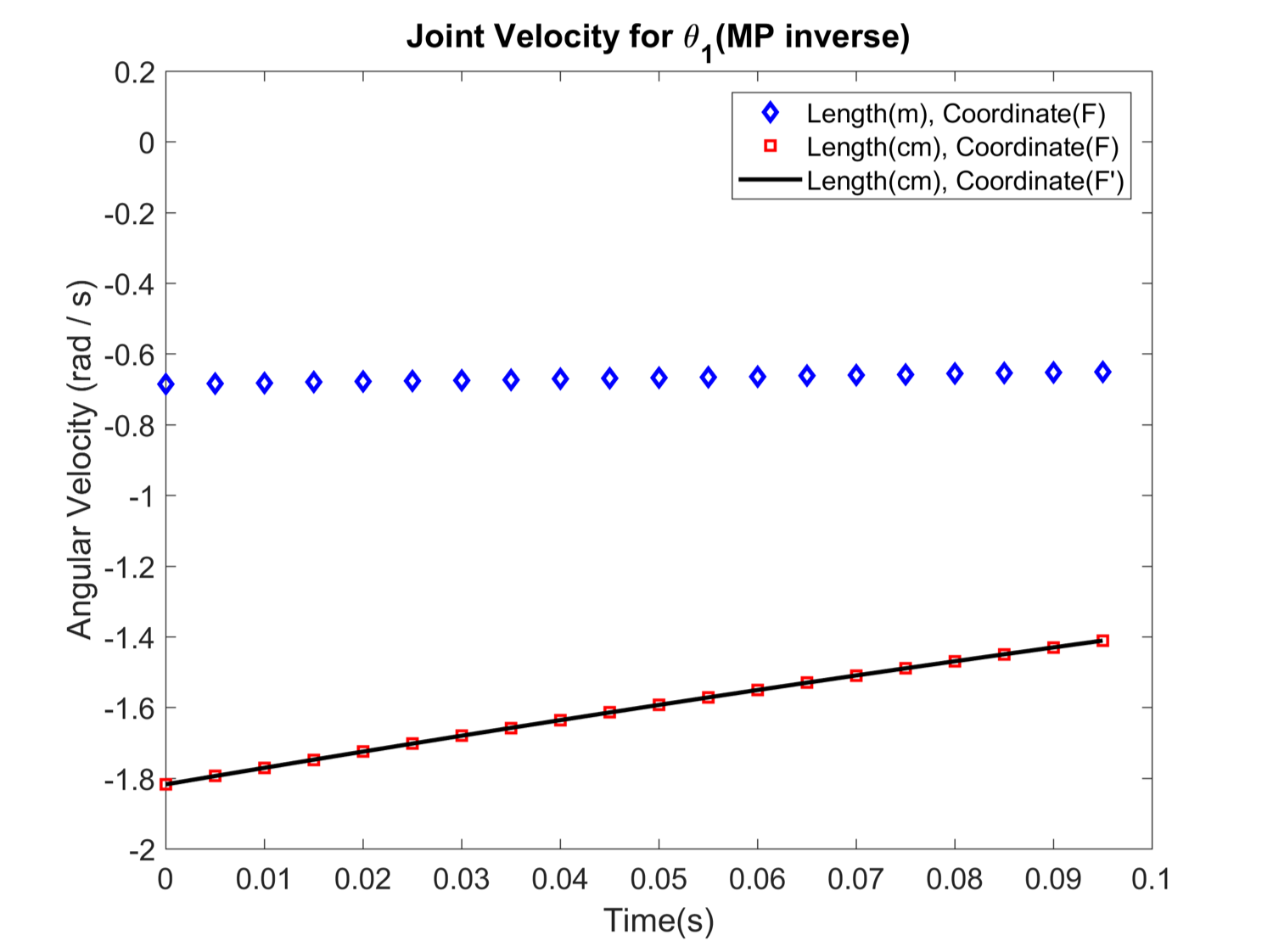}%
	\label{f6-1}}
	\hfil
	\subfloat[]{\includegraphics[width=3.3in]{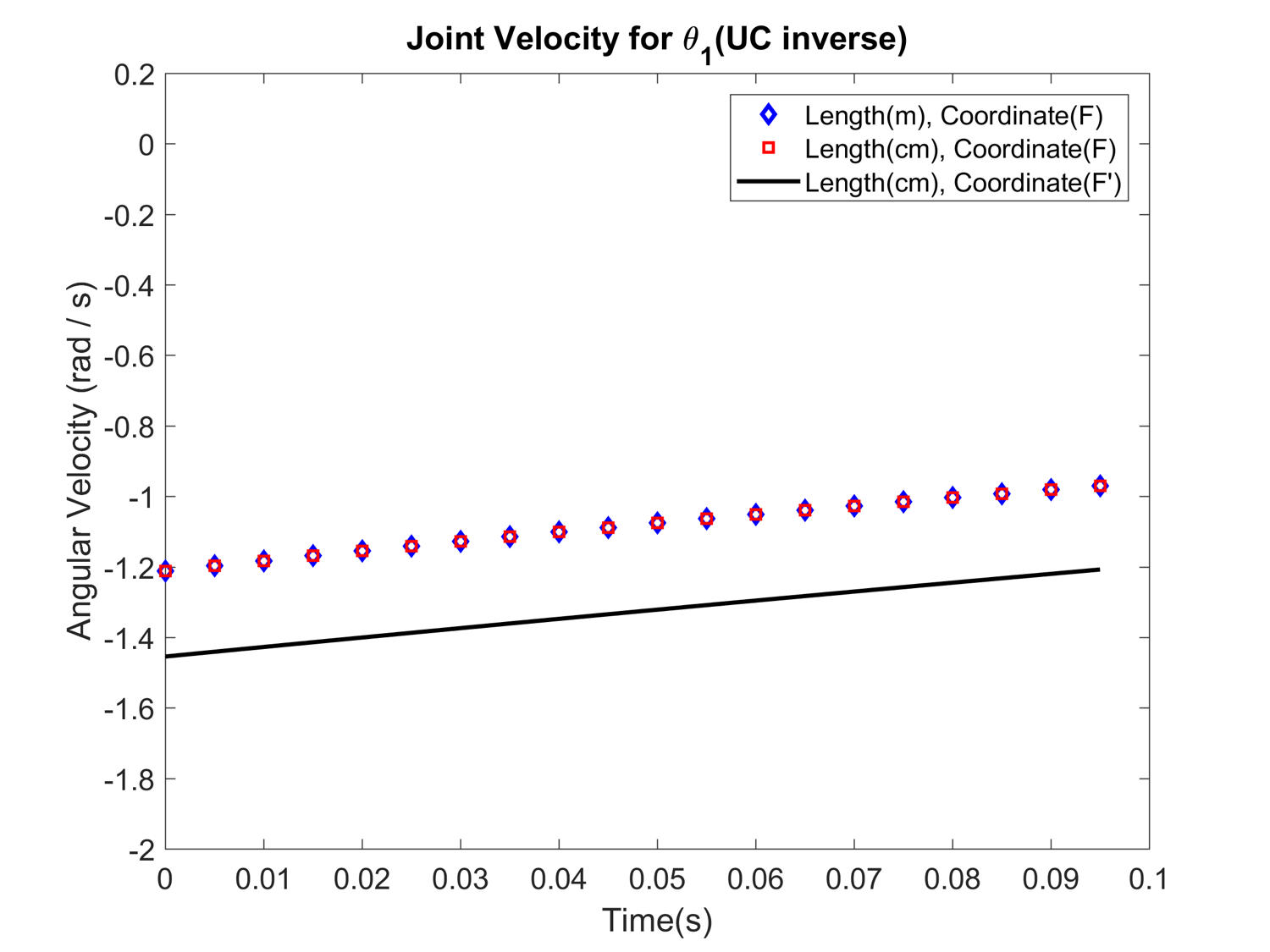}%
	\label{f6-2}}
	\hfil
	\subfloat[]{\includegraphics[width=3.3in]{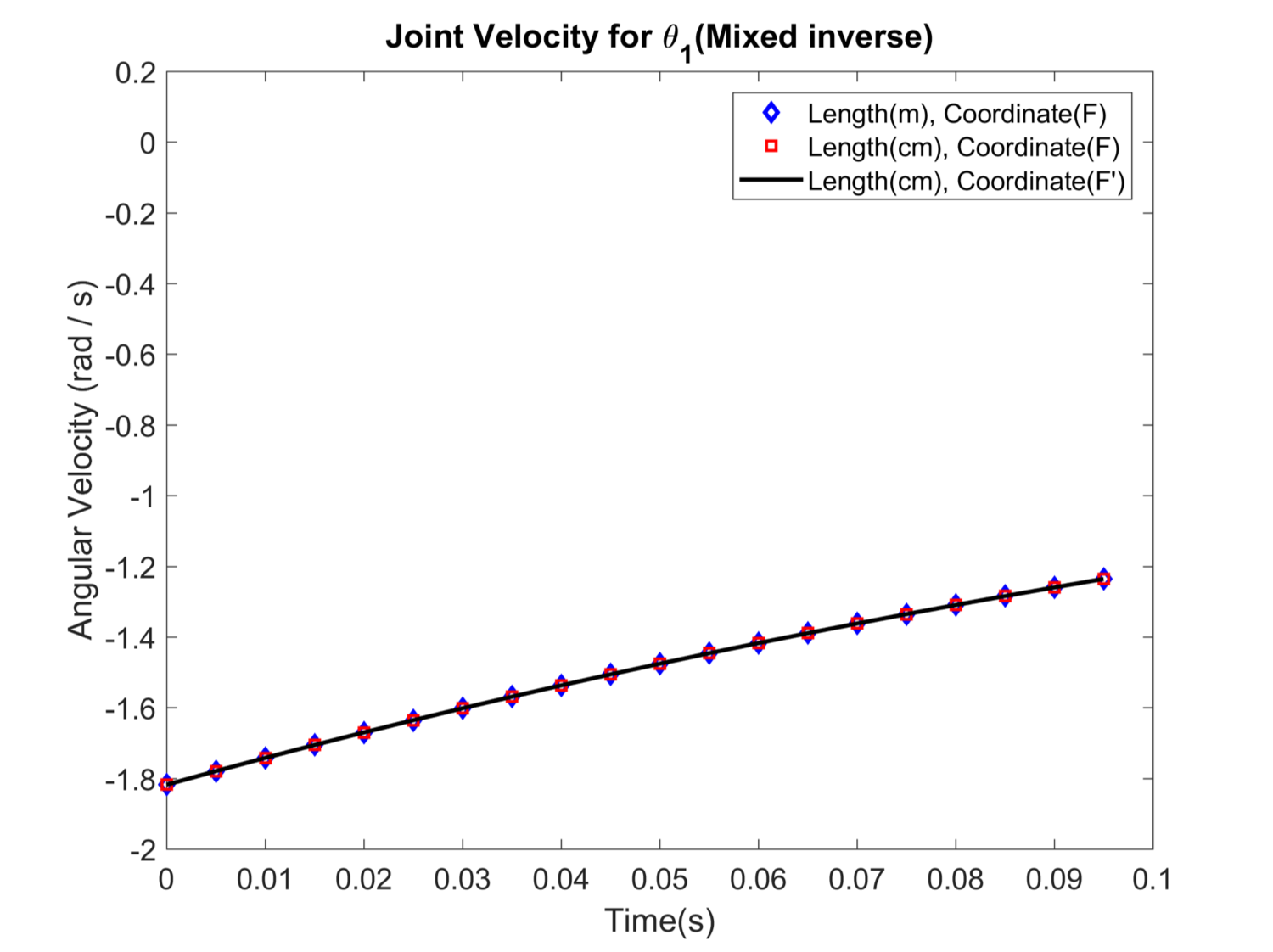}%
	\label{f6-3}}
	\caption{Joint velocity for $\theta_1$ solved with different generalized inverse approaches. (a) MP inverse. (b) UC inverse. (c) mixed inverse. Only the mixed inverse yields the same results over the $0.1s$ for the transformations in all three cases.}	 
	\label{f6}
\end{figure}

A transient simulation was performed for $0.1s$ to further observe the full control process. Figure \ref{f6}(a) displays variation of $\theta_1$ for the three approaches. It shows that the angular velocity calculated over time by the MP inverse is not affected by a rotation of the coordinate frame from $F$ to $F'$ but is affected by a change of the length unit from meters to centimeters; and the reverse is true for the UC inverse. By contrast, the angular velocity from the mixed inverse is identical over time in all cases. 

In summary, for this system involving variables with different consistency requirements the mixed inverse yields reliable control while the alternatives do not. This 
demonstrates the necessity of using the appropriate inverse to satisfy all applicable consistency requirements.  
\begin{center}
\begingroup
\renewcommand*{\arraystretch}{1.2}
\begin{table}
	\caption[]{Joint velocities $\dot q = [\dot \theta_1(rad/s), \dot l(m/s), \dot x_1(m/s), \dot y_1(m/s), 	\dot z_1(m/s)]^T$, solved with MP inverse, UC inverse, and Mixed inverse approaches.}
	\centering
	\begin{tabular}{|r||l|l|l|l|}
	\hline
	 & Length unit($m$) & Length unit($cm$) & Length unit($cm$) \\ 
	 & Coordinate($F$) & Coordinate($F$) & Coordinate($F'$) \\ 
	\hline
	MP inv & 
	$\dot q =
	\begin{bmatrix}
	-0.6854 \\
	0.8536 \\
	1.1963 \\
	-0.0498 \\
	-0.3964
	\end{bmatrix}
	$  & 
	$\dot q =
	\begin{bmatrix}
	-1.8179 \\
	0.8536 \\
	0.5734 \\
	-0.5731 \\
	-0.3964
	\end{bmatrix}
	$ & 
	$\dot q =
	\begin{bmatrix}
	-1.8179 \\
	0.8536 \\
	0.5734 \\
	-0.5731 \\
	-0.3964
	\end{bmatrix}
	$ \\
	\hline
	UC inv & 
	$\dot q =
	\begin{bmatrix}
	-1.2121 \\
	1.3536 \\
	0.6566 \\
	-0.0101 \\
	-0.0429
	\end{bmatrix}
	$  & 
	$\dot q =
	\begin{bmatrix}
	-1.2121 \\
	1.3536 \\
	0.6566 \\
	-0.0101 \\
	-0.0429
	\end{bmatrix}
	$ & 
	$\dot q =
	\begin{bmatrix}
	-1.4545 \\
	1.5690 \\
	 0.3676 \\
	-0.1943 \\
	-0.1095
	\end{bmatrix}
	$ \\
	\hline
	Mixed inv & 
	$\dot q =
	\begin{bmatrix}
	-1.8182 \\
	2.7071 \\
	-0.3536 \\
	-0.3536 \\
	0.9142
	\end{bmatrix}
	$  & 
	$\dot q =
	\begin{bmatrix}
	-1.8182 \\
	2.7071 \\
	-0.3536 \\
	-0.3536 \\
	0.9142
	\end{bmatrix}
	$ & 
	$\dot q =
	\begin{bmatrix}
	-1.8182 \\
	2.7071 \\
	-0.3536 \\
	-0.3536 \\
	0.9142
	\end{bmatrix}
	$ \\
	\hline
	\end{tabular}
	\label{t2} 
\end{table}
\endgroup
\end{center}

\section{Discussion}

In this paper we discussed the fact that different generalized inverses have different properties and that the correct one must be chosen based on the properties needed for the problem at hand. For example, the MP inverse is appropriate if the application requires the behavior of the system to be invariant with respect to rotations of the coordinate frame. We then described that many problems require the behavior of the system to be invariant with respect to the choice of units used for key parameters. For example, the behavior should be the same regardless of whether the parameters are defined in metric units or imperial units as long as all parameters are defined consistently with whatever choice of units is made. We emphasized that the Moore-Penrose (MP) pseudoinverse does not satisfy this requirement. 

We demonstrated what can happen if the wrong generalized inverse is used by examining the MP inverse in an example of a robotic arm in which consistency with respect to units is required. In that example it was shown that the MP inverse produced different behaviors depending on the choice of units, and as a result it produced erratic and unpredictable behaviors. This is because the MP inverse provides consistency with respect to rotations rather than changes of units. We described that methods are known to tailor such problems by hand so that unit consistency can be maintained, but those methods are complicated and thus are more likely to introduce accidental errors. We explained that what is truly needed to solve such problems is a generalized inverse that is unit consistent. Such an inverse, called the UC inverse, has recently been developed, and we showed that simply replacing the MP inverse with the UC inverse does in fact eliminate all dependencies on the choice of units. Use of the UC inverse led our example system to exhibit stable behavior in all cases.

Lastly we demonstrated that the UC and MP inverses can be combined to provide consistency for
systems defined with a mix of variables, some of which demand unit consistency while others
require rotational consistency.

\begin{appendices}
\section{UC Inverse and the Kronecker Product}

The Kronecker product is often used for the mathematical representation of a complex system 
in terms of simpler subsystems~\cite{lei18}. In this appendix we show that the UC inverse satisfies the
same useful properties as the MP inverse with respect to the Kronecker product.

The Kronecker product is a non-commutative tensor operator, usually denoted as $\otimes$, which takes an $m\times n$ matrix and a $p\times q$ matrix 
and constructs a composition of the two matrices to produce a higher-dimensional $mp\times nq$ matrix. 
The definition of the Kronecker product of $A_{m \times n}$ and $B_{p \times q}$ is 
$$
A \otimes B = 
\begin{bmatrix}
	a_{1, 1}B & a_{1, 2}B & \cdots & a_{1, n}B \\
	a_{2, 1}B & a_{2, 2}B & \cdots & a_{2, n}B \\
	\vdots & \vdots &\ddots & \vdots \\
	a_{m, 1}B & a_{m, 2}B & \cdots & a_{m, n}B \\
\end{bmatrix}
$$
Thus each $a_{i, j}B$ is a of  $p \times q$ matrix. For example, assuming $A$ is $2\times 2$ and $B$ is $3\times 2$ (i.e., $m = n = 2$ and $p = 3, n = 2$) then 
$$
A \otimes B = 
\begin{bmatrix}
	a_{1, 1}b_{1, 1}  & a_{1, 1}b_{1, 2} &  a_{1, 2}b_{1, 1}  & a_{1, 2}b_{1, 2} \\
	a_{1, 1}b_{2, 1}  & a_{1, 1}b_{2, 2} &  a_{1, 2}b_{2, 1}  & a_{1, 2}b_{2, 2} \\
	a_{1, 1}b_{3, 1}  & a_{1, 1}b_{3, 2} &  a_{1, 2}b_{3, 1}  & a_{1, 2}b_{3, 2} \\
	a_{2, 1}b_{1, 1}  & a_{2, 1}b_{1, 2} &  a_{2, 2}b_{1, 1}  & a_{2, 2}b_{1, 2} \\
	a_{2, 1}b_{2, 1}  & a_{2, 1}b_{2, 2} &  a_{2, 2}b_{2, 1}  & a_{2, 2}b_{2, 2} \\
	a_{2, 1}b_{3, 1}  & a_{2, 1}b_{3, 2} &  a_{2, 2}b_{3, 1}  & a_{2, 2}b_{3, 2} \\
\end{bmatrix}
$$
The Kronecker product is important in engineering design because it can be used to elegantly and
efficiently represent complex systems as compositions of simpler subsystems. It finds
applications in control systems, signal processing, image processing, semidefinite programming, and 
quantum computing\cite{van00, graham81, lei18, rachel18, boots13, saba17, daniel14}. 
It is bilinear and associative:
\begin{eqnarray}
	 A \otimes (B + C) = A \otimes B + A \otimes C \label{4-1}\\
	 (A +B )\otimes C =A \otimes C +B \otimes C \label{4-2}\\
	 (kA )\otimes B =A \otimes (kB)=k(A \otimes B) \label{4-3}\\
	 (A \otimes B )\otimes C =A \otimes (B \otimes C ) \label{4-4}
\end{eqnarray} 
and it satisfies the following with respect to the transpose (and conjugate-transpose) operator
\begin{equation} \label{4-5}
	(A \otimes B)^T = A^T \otimes B^T
\end{equation}
For matrices $A, B, C$ and $D$ for which the products $AC$ and $BD$ are valid,
the following mixed-product property (so-called because it involves both
standard matrix multiplication and the Kronecker product) can be shown to hold:
\begin{equation} \label{4-6}
	 (A \otimes B)(C \otimes D) = (AC) \otimes (BD) 
\end{equation}
If matrices $A$ and $B$ are orthogonal then
$A \otimes B$ is also orthogonal:
\begin{equation} \label{4-7}
	  (A \otimes B)^{T} (A \otimes B) = I
\end{equation}
It is also the case that
\begin{equation} \label{4-8}
	  (A \otimes B)^{-1} = A^{-1} \otimes B^{-1}
\end{equation}
and more generally
\begin{equation} \label{4-9}
	(A_{1} \otimes A_{2} \otimes \cdots \otimes A_{n})^{-1} = A_{1}^{-1} \otimes A_{2}^{-1} \otimes \cdots \otimes A_{n}^{-1}
\end{equation}
The last two properties are necessary to construct a matrix to transform
from one Kronecker-constructed matrix to another Kronecker-constructed matrix, which is 
required for performing controls of the kinds of robotic and mechanical
systems of interest in this thesis, but we also require the ability
to apply generalized matrix inverses in the case of singular matrices.
It has been proven that the MP inverse satisfies \cite{amy04}: 
\begin{equation} \label{4-10}
	  (A \otimes B)^{-P} = A^{-P} \otimes B^{-P}
\end{equation}
and more generally:
\begin{equation} \label{4-11}
	(A_{1} \otimes A_{2} \otimes \cdots \otimes A_{n})^{-P} = A_{1}^{-P} \otimes A_{2}^{-P} \otimes \cdots \otimes A_{n}^{-P}
\end{equation}
but it has not yet been established that these two results also hold for the UC inverse. 
They must be proven in order to show that the UC inverse can be used for general control systems represented 
using Kronecker products. We begin by proving that the Kronecker product of two diagonal matrices is also diagonal. 
Given diagonal matrices $A$ and $B$
$$
A = 
\begin{bmatrix}
	a_{1, 1} & 0 & \cdots & 0 \\
	0 & a_{2, 2} & \cdots & 0 \\
	\vdots & \vdots &\ddots & \vdots \\
	0 & 0 & \cdots & a_{m_A, m_A}
\end{bmatrix}, 
$$
$$
B = 
\begin{bmatrix}
	b_{1, 1} & 0 & 0 & \cdots & 0 \\
	0 & b_{2, 2} & 0 & \cdots & 0 \\
	\vdots & \vdots & \vdots &\ddots & \vdots \\
	0 & 0 & 0 & \cdots & b_{m_B, m_B}
\end{bmatrix}
$$
the Kronecker product
$$
C = A \otimes B = 
\begin{bmatrix}
	a_{1, 1}B & 0 & \cdots & 0 \\
	0 & a_{2, 2}B & \cdots & 0 \\
	\vdots & \vdots &\ddots & \vdots \\
	0 & 0 & \cdots & a_{m_A, m_A}B
\end{bmatrix}
$$
can be seen to also be diagonal because every nonzero element of $A$ is at position $[i, i]$ ($1 \leq i \leq m$), 
every nonzero element of $B$ is at position $[j, j]$ ($1 \leq j \leq p$), and every nonzero element of $C$ is at 
position $[i (p - 1) + j, i (p - 1) + j]$. Or more simply, every diagonal block of $C$ is a diagonal matrix
and therefore $C$ must be a diagonal matrix.

With these basic properties in mind, we have the prerequisites to prove the UC inverse property. First consider the base case, 
\begin{equation} \label{4-12}
	A^{-U}_{1} \otimes A^{-U}_{2} = (A_{1} \otimes A_{2})^{-U}
\end{equation}
Recall the decomposition of equation \ref{3-21} for any matrix $A$, where for notational clarity we now use $D$ and $E$ instead
of $D_A$ and $E_A$:
\begin{equation} \label{4-13}
	A = D \cdot S \cdot E
\end{equation}
and the UC generalized inverse of $A$ is defined as:
\begin{equation} \label{4-14}
	A^{-U} = E^{-1} \cdot S^{-P} \cdot D^{-1}
\end{equation}
Applying this to the left hand side of equation \ref{4-12} gives
\begin{eqnarray}
	&&A^{-U}_{1} \otimes A^{-U}_{2} \nonumber \\
	&=& (E_{1}^{-1} \cdot S_{1}^{-P} \cdot D_{1}^{-1}) \otimes (E_{2}^{-1} \cdot S_{2}^{-P} \cdot D_{2}^{-1}) \label{4-15} \\
	&=& (E^{-1}_{1}  \otimes E^{-1}_{2} )(S^{-P}_{1}  \otimes S^{-P}_{2})( D^{-1}_{1}  \otimes D^{-1}_{2}) \label{4-16} \\
	&=& (E_{1}  \otimes E_{2} )^{-1}(S^{-P}_{1}  \otimes S^{-P}_{2})( D_{1}  \otimes D_{2})^{-1} \label{4-17}
\end{eqnarray}
Using the fact that  $S^{-P}_{1}  \otimes S^{-P}_{2} = (S_{1}  \otimes S_{2})^{-P}$ gives
\begin{equation} \label{4-18}
	A^{-U}_{1} \otimes A^{-U}_{2} = (E_{1}  \otimes E_{2} )^{-1}(S_{1}  \otimes S_{2})^{-P}( D_{1}  \otimes D_{2})^{-1}
\end{equation}
and the right-hand side of equation \ref{4-12} is 
\begin{eqnarray}
	&(A_{1} \otimes A_{2})^{-U} = [(D_{1} \cdot S_{1} \cdot E_{1}) \otimes (D_{2} \cdot S_{2} \cdot E_{2})]^{-U} \label{4-19} \\
	&= [(D_{1} \otimes D_{2})(S_{1} \otimes S_{2})( E_{1} \otimes E_{2})]^{-U} \label{4-20}
\end{eqnarray}
where $D_{1} \otimes D_{2}$ and $E_{1} \otimes E_{2}$ are diagonal matrices. In order to apply equation \ref{4-14}, the rows and 
columns of $S_{1} \otimes S_{2}$ must satisfy the $\pm 1$ product constraint. Let $S_{1}$ and $S_{2}$ be represented as
$$
S_{1} = 
\begin{bmatrix}
	a_{1, 1} & a_{1, 2} & \cdots & a_{1, n_1} \\
	a_{2, 1} & a_{2, 2} & \cdots & a_{2, n_1} \\
	\vdots & \vdots &\ddots & \vdots \\
	a_{m_1, 1} & a_{m_1, 2} & \cdots & a_{m_1, n_1} \\
\end{bmatrix}
$$
$$
S_{2} = 
\begin{bmatrix}
	b_{1, 1} & b_{1, 2} & \cdots & b_{1, n_2} \\
	b_{2, 1} & b_{2, 2} & \cdots & b_{2, n_2} \\
	\vdots & \vdots &\ddots & \vdots \\
	b_{m_2, 1} & b_{m_2, 2} & \cdots & b_{m_2, n_2} \\
\end{bmatrix} 
$$
where for any $1 \leq i_1 \leq m_1$, $1 \leq j_1 \leq n_1$, $1 \leq i_2 \leq m_2$, $1 \leq j_2 \leq m_2$, the matrix elements 
of $S_{1}$ and $S_{2}$ have the following property: \\
\begin{eqnarray}
\prod a_{i_1, k} = \pm 1   \qquad  (a_{i_1, k} \neq 0) \label{4-21} \\
\prod a_{k, j_1} = \pm 1    \qquad (a_{k, j_1} \neq 0) \label{4-22}\\
\prod b_{i_2, k} = \pm 1    \qquad (b_{i_2, k} \neq 0)  \label{4-23}\\
\prod b_{k, j_2} = \pm 1    \qquad (b_{k, j_2} \neq 0)  \label{4-24}
\end{eqnarray}
For every row $i_1(m_2 - 1) + i_2$ of
$$
S_{1} \otimes S_{2} = 
\begin{bmatrix}
	a_{1, 1}S_{2} & a_{1, 2}S_{2} & \cdots & a_{1, n_1}S_{2} \\
	a_{2, 1}S_{2}& a_{2, 2}S_{2} & \cdots & a_{2, n_1}S_{2} \\
	\vdots & \vdots &\ddots & \vdots \\
	a_{m_1, 1}S_{2} & a_{m_1, 2}S_{2} & \cdots & a_{m_1, n_1}S_{2} \\
\end{bmatrix}
$$
the product of its nonzero elements is (when $a_{i_1, k_1} \neq 0, b_{i_2, k_2} \neq 0$)
\begin{eqnarray}
	&&\prod (a_{i_1, k_1}(\prod b_{i_2, k_2}) )       \label{4-25} \\
	& = &\prod (a_{i_1, k_1}(\pm 1) )  \label{4-26}\\
	& = &\prod (a_{i_1, k_1})  \prod (\pm 1)  \label{4-27}\\
	& = & \prod (\pm 1)  \label{4-28}\\
	& = & \pm 1.  \label{4-29}
\end{eqnarray}
The same holds analogously for every column of $S_{1} \otimes S_{2}$, so equation \ref{4-14} can be applied to equation \ref{4-20} to obtain
\begin{eqnarray}
  (D_{1} \otimes D_{2})(S_{1} \otimes S_{2})( E_{1} \otimes E_{2})]^{-U}  \nonumber \\
	~ = ~ (E_{1}  \otimes E_{2} )^{-1}(S_{1}  \otimes S_{2})^{-P}( D_{1}  \otimes D_{2})^{-1} \label{4-30}
\end{eqnarray}
and from equation \ref{4-18} the following theorem can be concluded: \\

\noindent \underline{Theorem 1}: $A^{-U}_{1} \otimes A^{-U}_{2} = (A_{1} \otimes A_{2})^{-U}$.\\

We now show that Theorem 1 can be used as the base case for a mathematical induction 
proof of the general case involving matrices $A_1 ... A_n$. Given 
\begin{equation} \label{4-32}
	(A_{1} \otimes A_{2} \otimes \cdots \otimes A_{n})^{-U} = A_{1}^{-U} \otimes A_{2}^{-U} \otimes \cdots \otimes A_{n}^{-U}
\end{equation}
it is required to show 
\begin{equation} \label{4-33}
	(A_{1} \otimes A_{2} \otimes \cdots \otimes A_{n+1})^{-U} = A_{1}^{-U} \otimes A_{2}^{-U} \otimes \cdots \otimes A_{n+1}^{-U}
\end{equation}
To simplify the equation, let $B = A_{1} \otimes A_{2} \otimes \cdots \otimes A_{n} $. Then
\begin{eqnarray} 
	(A_{1} \otimes A_{2} \otimes \cdots \otimes A_{n+1})^{-U} = (B \otimes A_{n+1})^{-U} \label{4-34}\\
	A_{1}^{-U} \otimes A_{2}^{-U} \otimes \cdots \otimes A_{n+1}^{-U} = B^{-U} \otimes A_{n+1}^{-U}. \label{4-35}
\end{eqnarray}
Applying Theorem 1 while Letting $A_{1} = B$ and $A_{2} = A_{n+1}$ we obtain
\begin{eqnarray} 
	(B \otimes A_{n+1})^{-U}  = B^{-U} \otimes A_{n+1}^{-U} \label{4-36}
\end{eqnarray}
from which we can expand to obtain the desired final result:\\

\noindent \underline{Theorem 2}: \\
$(A_{1} \otimes A_{2} \otimes \cdots \otimes A_{n+1})^{-U} = A_{1}^{-U} \otimes A_{2}^{-U} \otimes \cdots \otimes A_{n+1}^{-U}$\\

Theorems 1 and 2 complete the set of UC inverse properties necessary to allow it to be used in place of 
the MP inverse in mechanical and robotic systems whenever unit consistency is required.

\section{Implementation of UC Inverse}

\noindent The following Octave/Matlab implementation is adapted from code in\!~\cite{uhlmann}.

\begin{verbatim}
function Ai = uinv(A)
    tolerance = 1e-22;    
    [m, n] = size(A);
    L = zeros(m, n);    M = ones(m, n);
    S = sign(A);   A = abs(A);
    idx = find(A > 0.0);
    L(idx) = log(A(idx));
    idx = setdiff(1 : numel(A), idx);
    L(idx) = 0; A(idx) = 0; M(idx) = 0;   
    r = sum(M, 2);   c = sum(M, 1);   
    u = zeros(m, 1); v = zeros(1, n);
    sumv = 0.0; prev = 1.0;
    while (abs(prev - sumv) > tolerance)
        idx = c > 0;
        p = sum(L(:, idx), 1) ./ c(idx);
        L(:, idx) = L(:, idx) - repmat(p, m, 1) .* M(:, idx);
        v(idx) = v(idx) - p;
        idx = r > 0;
        p = sum(L(idx, :), 2) ./ r(idx);
        L(idx, :) = L(idx, :) - repmat(p, 1, n) .* M(idx, :);
        u(idx) = u(idx) - p;
        prev = sumv;
        sumv = var(sum(L));
    end    
    Ai = pinv(exp(L) .* S) .* (exp(u) * exp(v))';
end
\end{verbatim} 

\end{appendices}

\vspace{-0.5in}
\begin{IEEEbiography}[{\includegraphics[width=1in,height=1.25in,clip,keepaspectratio]{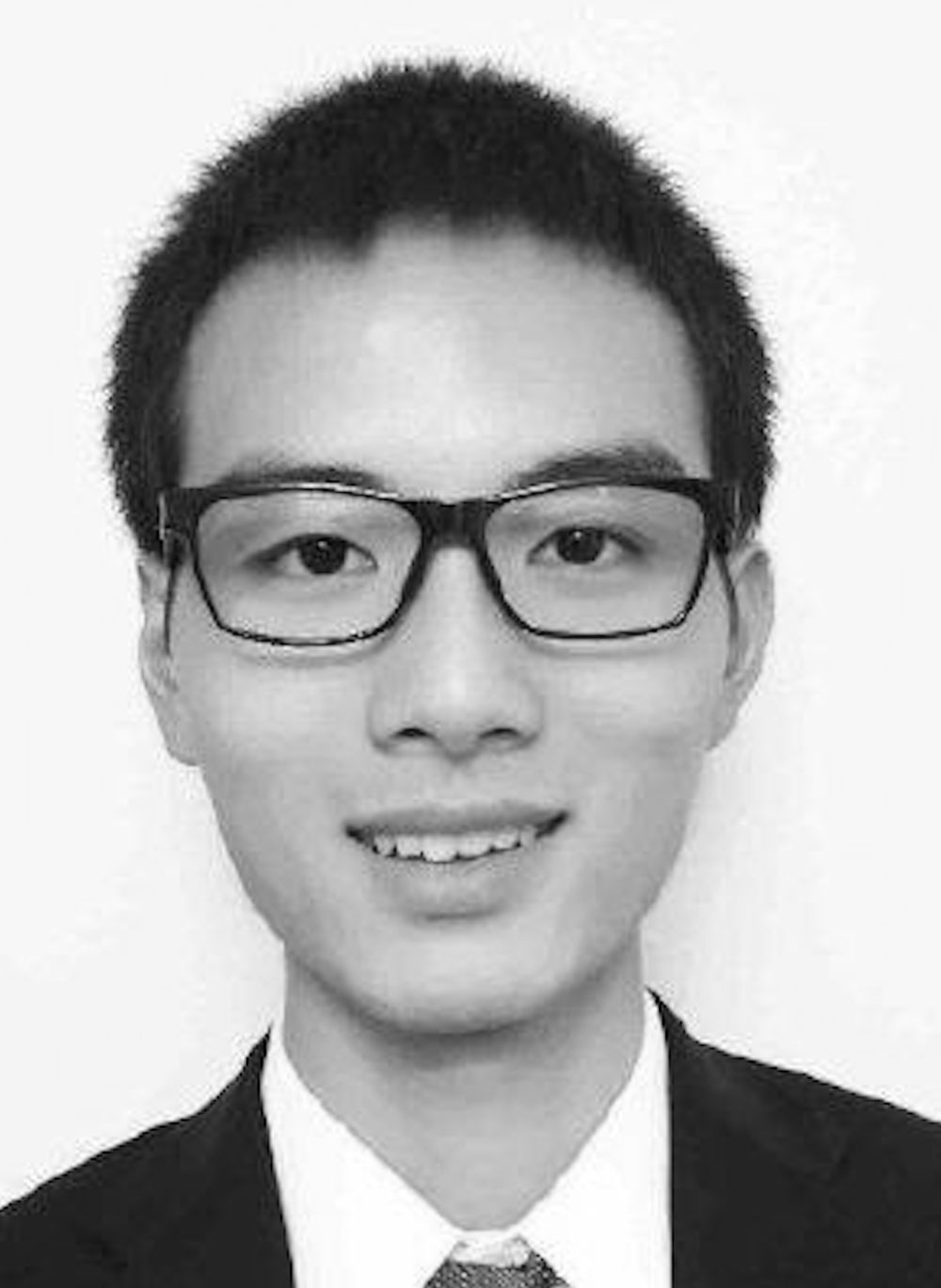}}]{Bo Zhang}
Bo Zhang received the B.S. degree in Mechanical Engineering from East China
University of Science and Technology, Shanghai, China in 2014 and is currently
working toward the M.S. degree in the Electrical Engineering and Computer
Science Department, University of Missouri-Columbia, MO.
His research interests include kinematic systems and computational analysis.
\end{IEEEbiography}
\vspace{-0.5in}
\begin{IEEEbiography}[{\includegraphics[width=1in,height=1.25in,clip,keepaspectratio]{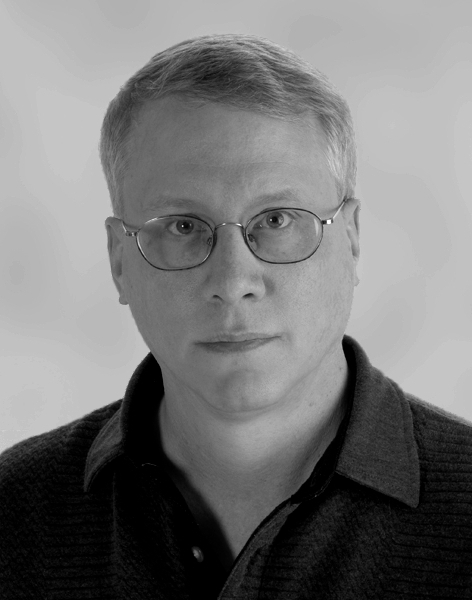}}]{Jeffrey Uhlmann}
Prof. Uhlmann is a faculty member of the Dept.\ of Electrical Engineering and Computer Science at the University of Missouri-Columbia. He received his doctorate in robotics from the University of Oxford, UK, in 1995 and was a research scientist for 13 years at the Naval Research Laboratory (NRL) in Washington, DC.
\end{IEEEbiography}

\end{document}